\documentclass[11pt,twoside,A4]{article}
\usepackage{times,fancyhdr}
\usepackage[dvips]{graphicx}
\usepackage{latexsym}
\usepackage[affil-it]{authblk}
\usepackage{mathptm}
\usepackage{amsmath}
\usepackage{amssymb}
\usepackage{floatrow}
\usepackage{units}
\usepackage{setspace}
\usepackage{hyperref}
\usepackage[font=small,labelfont=bf]{caption}
\usepackage[top=4cm, bottom=4cm, left=3.5cm, right=3.5cm]{geometry}

\def\beq{\begin{equation}}
\def\eeq{\end{equation}}
\def\beqn{\begin{eqnarray}}
\def\eeqn{\end{eqnarray}}

\begin{document}

\title{Born's rule from almost nothing}
\author{Sabine Hossenfelder}
\affil{Frankfurt Institute for Advanced Studies\\
Ruth-Moufang-Str. 1,
D-60438 Frankfurt am Main, Germany
}
\date{}
\maketitle
\vspace*{-1cm}

\begin{abstract}
We here put forward a simple argument for Born's rule based on the requirement that the probability distribution should not be a function of the number of degrees of freedom.
\end{abstract} 

\section{Introduction}

Quantum mechanics does not make definite predictions but only predicts probabilities for measurement outcomes. One calculates these probabilities
from the wave-function using Born's rule. In axiomatic formulations of quantum mechanics, Born's rule is usually added as an
axiom on its own right. However, it seems the kind of assumption that should not require a postulate, but that should instead follow from
the physical properties of the theory. 

Since the early days of quantum mechanics, there have thus been many attempts to derive 
Born's rule from other assumptions: non-contextuality \cite{Gleason}, decision theory \cite{Deutsch}, non-triviality \cite{Aaronson}, and the number of degrees of freedom in composite systems \cite{Hardy,Masanes:2010tt,Brukner}. The argument discussed here is most similar to the ones for many worlds presented in \cite{Vaidman, Carroll:2014mea} and the one using environment-assisted invariance \cite{Zurek}. However, as will become clear shortly, the ontological baggage of these arguments is unnecessary. 

\section{Argument}

To get started, let $|\Psi \rangle$ and $|\Phi \rangle$ be elements of a vector space over the complex numbers which denote points on the complex sphere of dimension $N \in {\mathbb{N}}^+$, i.e. $|\langle \Psi | \Psi \rangle| =1$ and $|\langle \Phi | \Phi \rangle| =1$.
\bigskip

{\bf Claim:} The only well-defined and consistent distribution for transition probabilities $P_N(|\Psi\rangle \to |\Phi\rangle)$ on the complex sphere of dimension $N$ which is continuous, independent of $N$, and invariant under unitary operations is $P_N(| \Psi \rangle \to |\Phi \rangle) = |\langle \Psi | \Phi \rangle|^2$. The continuity assumption is unnecessary if one restricts the original space to states of norm $K/N$ or, correspondingly, to rational-valued probabilities as a frequentist interpretation would suggest. 
\bigskip

\noindent By $P_N$ being a well-defined probability distribution we mean
\beqn
P_N(|\Psi\rangle \to |\Phi\rangle) \in \mathbb{R} ~\wedge~ 0 \leq P_N(|\Psi\rangle \to |\Phi\rangle) \leq 1 ~~~~\forall~ |\Psi \rangle, |\Phi \rangle ~. \label{welldef}
\eeqn
By $P_N$ being consistent, we mean
\beqn
&& \sum_{i=1}^N P_N(|\Psi\rangle \to |\Psi^i\rangle) = 1 ~~~~ \forall~ |\Psi \rangle~,~ \langle \Psi^i |\Psi^j\rangle = \delta^{ij} ~, \label{norm} \\
&& P_N(|\Psi^i\rangle \to |\Psi^j\rangle) = \delta^{ij} ~~~~ {\mbox{for}}~~~~ \langle \Psi^i |\Psi^j\rangle = \delta^{ij} \label{orth}~.
\eeqn

{\bf Proof}: Since $P_N$ is invariant under unitary operations, transition probabilities can only be functions of scalar products, ie 
$P_N(|\Psi \rangle \to |\Phi \rangle)  = P_N(\langle \Psi |\Phi \rangle) ~ \forall ~ |\Psi \rangle, |\Phi \rangle$. This means that $P_N$ is actually a map from the complex unit sphere of dimension 1 to the interval $[ 0,1] \in {\mathbb{R}}$. It follows from (\ref{orth}) that $P(\langle \Psi | \Phi \rangle) = 0$ whenever $\langle \Psi | \Phi \rangle =0$, because if $|\Psi \rangle$ and $|\Phi \rangle$ are orthogonal, one could construct a basis of which they are elements. This means $P(0) = 0$. That $P_N$ is independent of $N$ then just means that this function is the same for all $N$, so we will from now on omit the index $N$. 

Let $|\Psi^i \rangle$, $i \in \{1... N \}$ be an arbitrary orthonormal basis and
\beqn
|\Psi^* \rangle := e^{{\rm i} \theta} \sqrt{\frac{1}{N}} \sum_{i=1}^N |\Psi^i \rangle~,
\eeqn
where $\theta$ is a real number in $[0,2 \pi [$. Because of (\ref{norm}) we then have
\beqn
\sum_{j=1}^N P (\langle \Psi^j | \Psi^* \rangle)  = N  P( e^{{\rm i} \theta}/\sqrt{N}) =  1~,
\eeqn
which means
\beqn
P( e^{{\rm i} \theta} /\sqrt{N}) = 1/N ~. \label{1/N}
\eeqn

Next we use a new basis, $|\widetilde \Psi^i \rangle$ with $i \in \{1..N \}$. With $K \in {\mathbb{N}}^+$, $K < N$ we define
\beqn
| \widetilde \Psi^j \rangle &:=& \sqrt{\frac{1}{K}} \sum_{l=1}^K \exp \left( -2 \pi {\rm i} \frac{(l-1) (j-1)}{K} \right) |\Psi^l \rangle  ~~~~{\mbox{for}}~1 \leq j \leq K~, \nonumber\\
| \widetilde \Psi^j \rangle &:=& |\Psi^j \rangle ~~~~{\mbox{for}}~K+1 \leq j \leq N~. \label{tildebasis}
\eeqn
The first $K$ basis-vectors are obviously orthogonal to the last $N-K$ and we already know that the last $N-K$ are orthonormal, so it remains to show that the first $K$ basis vectors are orthogonal. For this we note that for $j,m \leq K$
\beqn
\langle \widetilde \Psi^j | \widetilde \Psi^m \rangle 
= \frac{1}{K} \sum_{l=1}^ K        
\exp \left( -2 \pi {\rm i} \frac{m-j}{K} \right)^{l-1} ~.
\eeqn
For $m=j$ the sum gives $=K$, so the norm is $=1$ as it should be. For $m \neq j$ the sum is a geometric series and we have
\beqn
\langle \widetilde \Psi^j | \widetilde \Psi^m \rangle = \frac{ 1- \exp \left( -2 \pi {\rm i} (m-j)\right)} {1- \exp \left( -2 \pi {\rm i} (m-j) /K \right)  } =0 ~.
\eeqn

The first basis vector $|\widetilde \Psi^1 \rangle$ is paralell to $|\Psi^* \rangle$ in the subspace spanned by the first $K$ basisvectors $|\Psi^i \rangle$, therefore the $|\widetilde \Psi^j \rangle$'s for $2\leq j \leq K$ are also orthogonal to $|\Psi^*\rangle$. With this, and using $P(0) = 0$, we get
\beqn
\sum_{j=1}^N P (\langle \widetilde \Psi^j | \Psi^* \rangle) = P( e^{{\rm i} \theta}  \sqrt{K/N} ) + \sum_{j=K+1}^N P (  e^{{\rm i} \theta}  /\sqrt{N}) = 1~.
\eeqn
And by using (\ref{1/N}) we arrive at
\beqn
P ( e^{{\rm i} \theta} \sqrt{K/N} ) = K/N ~~~~\forall~ K,N \in {\mathbb{N^+}}, K \leq N, \theta \in [0, 2 \pi [~. \label{this}
\eeqn
If one was dealing with probabilitites that only took on rational values, then one can stop here. For real-valued probabilities one merely notes that
the only continuous function on the complex unit sphere with property (\ref{this}) is $P(x) = |x|^2$, hence
\beqn
P_N ( |\Psi \rangle \to |\Phi \rangle) = |\langle \Psi | \Phi \rangle|^2 ~~~~~~~~~~ \Box
\eeqn

\section{Discussion}

The verbal summary of this argument is that since the probability distribution has to be invariant under unitary operations, we can permute all basis elements, which means that the probabilities of a symmetric state like $|\Psi^*\rangle$ have to be evenly distributed, which fixes them. Then we can combine any $K$-dimensional subset of these basis elements to one whose probability is just the sum of probabilities assigned to the previous $K$ basis elements. Again we can do that in arbitary permutations and this fixes the probabilities for all $K/N$. The assumption that the distribution is independent of $N$ is essential; it is what allows us to fill in the sphere densely. 

\section*{Acknowledgements}

I thank Scott Aaronson, Sandro Donadi, and Tim Palmer for helpful feedback. This research was supported by the German Research Foundation (DFG) and the Franklin Fetzer Fund.

\end{document}